\documentstyle[12pt,floats,prl,aps,epsf,epsfig]{revtex}
\begin{document}
     \title{
     \large\bf
     An $\dot{f}(f)$-frequency dynamics
               algorithm for gravitational waves}
     \author{Maurice H.~P.~M. van Putten 
\footnote{mvp@schauder.mit.edu}
     and Abhinanda Sarkar
\footnote{sabhinan@in.ibm.com}
}
     \address{Massachusetts Institute of Technology,
	      Cambridge, MA 02139}
	       
     \maketitle
	
\baselineskip16pt
\mbox{}\\
\centerline{\bf Abstract}
\mbox{}\\
{\textsc{\bf
Coalescence of low mass
compact binaries of neutron stars and black holes
are primary burst sources for LIGO and VIRGO.
Of importance in the early stages of observations will
be the classification of candidate detections by source-type.
The diversity in source parameters and serendipity in any
new window of observations suggest to consider model-independent
detection algorithms.
Here a frequency dynamics algorithm is described which extracts
a trajectory in the $\dot{f}(f)$-plane from the noisy signal. 
The algorithm is studied in simulated binary coalescence.
Robust results are obtained with experimental noise data.
Experiments show the method to be superior to matched filtering 
in the presence of model imperfections.} 
\section{Introduction}

This millennium will mark the beginning of 
gravitational wave astronomy with the broad band
observatories LIGO\cite{AB:92} and VIRGO\cite{BR:92}.
This opens up new opportunities to probe strongly gravitating processes
in the coalescence of binaries 
and the birth of 
neutron stars or black holes\cite{CU:93,TH:95,SC:99}.
The late stages of spiral infall of
a neutron star may show interesting new physics, for example, in association
with tidal break-up around a companion black hole,
where the onset will provide
constraints on the density of neutron star matter.
There may also be surprises. Advanced LIGO is expected to see events
up to a Gpc\cite{TH:95,SC:99},
and might detect correlations or anti-correlations
of gravitational radiation
with cosmological $\gamma$-ray bursts.
The early stages of observations will be focused on verification
of potential gravitational wave signals and
identification of source by type.
In view of the diversity of source parameters and 
serendipity in any new window of observations,
it is of interest to consider model-independent detection algorithms.
These algorithms
are expected to compare well with matched filtering, anticipating
the inevitable model imperfections in both the wave-forms and the
instrumentation noise.

Early evolution of a binary of two stars of mass $M_1$ and $M_2$
with orbital period $P$ is well-described by quasi-Newtonian evolution 
in the point-particle limit, 
whose luminosity ${\cal L}_{GW}$ in gravitational wave emissions
in the theory of general relativity is given by the Peter and Mathews'
formula\cite{PM:63} 
\begin{eqnarray}
{\cal L}_{GW}\sim\frac{32}{5}~(\omega{\cal M})^{10/3} F(e)
\end{eqnarray}
in geometrized units with $G=c=1$,
where $\omega=2\pi/P$ denotes the orbital angular velocity, 
$F(e)$ accounts for an orbital ellipticity $e$
and ${\cal M}
=(M_1M_2)^{3/5}/(M_1+M_2)^{1/5}$
 denotes the chirp mass. 
This is accompanied
by a gradual rise 
in the orbital frequency $f_{orb}=1/P$:
\begin{eqnarray}
\dot{f}_{orb}=\frac{96}{5}(2\pi)^{8/3}{\cal M}^{5/3}
        f_{orb}^{{11}/{3}}.
\label{EQN_GW}
\end{eqnarray}
The predicted luminosity
of $3.2\times 10^{33}$ ergs/s agrees with the
observed rate of change 
$2.40\times 10^{-12}$ in the
orbital period of about eight hours in
the Hulse-Taylor binary system with neutron star masses
$M_1=1.42M_\odot$ and $M_2=1.4M_\odot$\cite{HT:xx}.
This strong frequency-luminosity relationship suggests to plot
the evolution of a candidate source as a 
trajectory in the $\dot{f}(f)$-plane
as the output of
a detection algorithm. 
In doing so, we focus on burst sources whose gravitational waves show
appreciable dynamics in their frequencies. This, in contrast to periodic
sources, e.g.: rapidly spinning
neutron stars\cite{SC:97,BI:98}, which we regard as a separate class of
sources in regards to detection algorithms.
Binary coalescence of compact objects, then, shows
an initial branch $\dot{f}\propto f^{11/3}$ indicative of 
the general relativistic relationship ${\cal L}_{GW}\propto f^{10/3}$.
These trajectories terminate on the $f$-axis at 
the quasi-normal mode frequency $f_{QNR}$ of a final
black hole. For the expected low mass black holes $M\le 10M_\odot$ in 
compact binaries, $f_{QNR}\sim 2800 (10M_\odot/M)$ is
beyond the (advanced) LIGO sensitivity range of about 
(20-1600Hz) 40-800Hz, however.
The transition which connects the chirp to the final black hole state
remains highly uncertain - here the gravitational radiation emission
process is most nonlinear and potentially most interesting.
For black hole-black hole coalescence, the transition from
chirp to a common horizon envelope state should be smooth, may be
very luminous\cite{TH:95} and more frequent\cite{PZ:99} than
neutron star-black hole coalescences\cite{PH:91,RN:91,PH:97}.
In neutron star-black hole coalescence, on the other hand, an
intermediate black hole-torus state is expected if the black hole
spins rapidly\cite{PA:91}. This state is expected to be
quiescent in its gravitational wave emissions, while luminous in 
emissions by the black hole in contact with the magnetic field of the
torus\cite{VP:2000}.
This predicts an anti-correlation between gravitational wave--emissions
and $\gamma$-ray bursts from this type of catastrophic events\cite{VP:99}.
Alternatively, gravitational radiation is expected in accretion induced 
collapse of a white dwarf into a neutron
star or black hole due to a bar mode instability\cite{US:92},
or the collapse of a young massive star\cite{WO:1993,PA:1998},
in which the relationship between $\dot{f}$ and $f$ is even 
more uncertain. If the collapse stalls with the formation of
a neutron star, we might witness a negative branch $\dot{f}<0$ due to
spin-down by gravitational wave emissions.
It becomes potentially useful, therefore, to plot
transient gravitational wave emissions as trajectories 
in the $\dot{f}(f)$-plane when
classifying candidate detections by source-type\cite{VP:99b}.
As gravitational wave-emissions are not expected to be 
significantly beamed, observations such as these by
advanced LIGO can obtain definite evidence
of $\gamma$--ray burst progenitors and their population statistics which 
are not otherwise available.

Here we describe an algorithm which enables accurate extraction
of the frequency dynamics in the
gravitational wave signal. The algorithm is based on counting
zero-crossings, wherefrom an
instantaneous frequency $f(t)$ and frequency rate of change
$\dot{f}(t)$ can be estimated as a function of observer's time $t$.
Here the proposed algorithm is tested using simulated binary  
coalescence with Gaussian noise and noise recorded on
the 40m LIGO test facility at Caltech.
The algorithm is studied on 
signals of the type
\begin{eqnarray}
X(t)=A(t)\cos(2\pi\Phi(t))
+\sigma\tilde{N}(t)
\end{eqnarray}
wth amplitude $A(t)=(1-t)^{-1/4}$ and phase 
\begin{eqnarray}
\Phi(t)=c\left[
\left(1-(1-t)^{5/8}\right)+
p\left(1-(1-t)^{3/8}\right)\right],
\end{eqnarray}
where $t\epsilon[0,1)$ is the time normalized to the time of coalescence
$t=1$, $c=1000$ and $p\epsilon[0,1]$ is a model parameter: 
$p=0$ in the Newtonian approximation and
$p=0.6038$ in the 1PN approximation when
$M_1=M_2$\cite{WA:77}.
The noise process $\tilde{N}(t)$
is normalized such that $\sigma$ controls
the signal-to-noise ratio.
All simulations are for
data strings of a total of $N=10,000$ samples 
at $t_n=n\Delta t$ ($n=1,2,\cdots N$).
The instantaneous frequency $f(t)$ is
given by $d\Phi(t)/dt$ and
$d^2\Phi(t)/dt^2$ denotes its
rate of change $\dot{f}(t)$ with respect to time.
The dimensionless rate of change of the period,
$f^{-2}\dot{f}$, expresses inversely the number of periods at a
frequency $f$, which is of the order of a few hundred in
the experiments reported below (final time $t\sim 0.97$).

\section{Description of the algorithm}

Consider the instantaneous frequency $f(t)$ as a finite time-series 
for $t\;\epsilon\;[a,b]$. We begin with
a partition of $[a,b]$ into $N$ adjacent subwindows 
(bins) $[s_{i-1},s_i]$ with centers at $t_i=(s_{i-1}+s_i)/2$.
Linear approximations
$f(t)\sim \alpha_i+\beta_i t$ ($s_{i-1}<t<s_i$) are then
obtained by estimates
of the coefficients $\alpha_i$ and $\beta_i$ from
the noisy data $X(t)$ 
in each subwindow $i$.
The number of periods in the $i$-th subwindow, for example, 
is $\int_{s_{i-1}}^{s_i}f(t)dt
=\alpha_i\delta+\beta_it_i\delta
+O(\delta^3)$,
where $\delta=s_i-s_{i-1}$.

Let $Z_S(s_{i-1},s_i)$ denote the number of zero-crossings of the true
signal $S(t)$ in the subwindow
$[s_{i-1},s_i]$. Then 
$Z_S(s_{i-1},s_i)$ differs from
$2\int_{s_{i-1}}^{s_i}f(t)dt$ by at most one. 
In view of $f(t)=d\Phi(t)/dt$,
our algorithm hinges on 
the approximation
\begin{eqnarray}
Z_S(s_{i-1},s_{i}) 
\sim 2\delta\alpha_i
    +2\delta\beta_it_i.
\end{eqnarray} 
The number of zero-crossings
in the noisy data
$\left(X(u_0),X(u_1),\cdots,X(u_k)\right)$ in
$[s_{i-1},s_i]$ 
will be denoted by $Z_X(s_{i-1},s_i)$, where
the $u_j$ denote the discrete sampling times
provided by the AD-converter (16 bit resolution,
10kHz).
The $X(u_j)$ have expectation values
$S(u_j)$ and 
variance $\sigma^2$. 
The 10kHz-sampling frequency is assumed to be 
sufficiently high such that the true signal agrees with the zero-crossings
of its linear interpolant based on the $S(u_j)$. In terms of the
indicator function, $I$, defined by $I(P)=1$ if $P$ is true and
$I(P)=0$ if $P$ is false, we have
\begin{eqnarray}
Z_S(s_{i-1},s_i)=\Sigma_1^k I[S(u_{j-1})S(u_j)<0],~~
Z_X(s_{i-1},s_i)=\Sigma_1^k I[X(u_{j-1})X(u_j)<0].
\end{eqnarray}
The probability $p_{j}$
that the observed signal $X(u_{j})$ and the its expectation (the true signal)
$S(u_{j})$ have the same sign can be easily computed. Suppose that 
$S(u_{j})>0$. Then, we have
\begin{eqnarray} 
P[X(u_{j})>0] = 
P[(X(u_{j})-S(u_{j}))/\sigma>-S(u_{j})/\sigma] = 1-F[-S(u_{j})/\sigma] 
= F[S(u_{j})/\sigma], 
\end{eqnarray}
where $F$ is the probability distribution of the noise process
$\tilde{N}(t)$, which is assumed to be symmetric about zero.
Similarly when $S(u_{j})<0$, we find $P[X(u_{j})<0] = F[-S(u_{j})/\sigma]$.  
It follows that $p_j=F[|S(u_j)|/\sigma]$, whereby
\begin{eqnarray}
P[I[X(u_{j-1})X(u_j)<0]\ne I[S(u_{j-1})S(u_j)<0)]]
=p_{j-1}(1-p_j)+p_j(1-p_{j-1}).
\end{eqnarray}
Consider, then, 
$R_i=\mbox{min}_j |S(u_j)|/\sigma$, 
as a measure for the 
signal-to-noise ratio in the subwindow 
$[s_{i-1},s_i]$. As $R_i\rightarrow\infty$, 
$\mbox{min}_j p_j\!\rightarrow\!1$. 
We then have for the expectation $E$ and the variance $V$
the limits
\begin{eqnarray}
\lim_{R_i\rightarrow\infty}
E[Z_X(s_{i-1},s_i)]=Z_S(s_{i-1},s_i),~~
V[Z_X(s_{i-1},s_i)]=0.
\end{eqnarray}
Hence, for large signal to noise ratios, we recover
$Z_X(s_{i-1},s_i)\sim 2\delta\alpha_i+2\delta\beta_it_i.$

Estimates $\alpha_i$ for
$f(t_i)$ and $\beta_i$ for
$\dot{f}(t_i)$ 
can be calculated by linear regression.
Define 
(as is usual in such regressions, see, e.g.
\cite{VR:1997})
the $L^2$ error
\begin{eqnarray}
Q(\alpha_i,\beta_i)
=\Sigma_1^N
\left(Z_{X}(s_{i-1},s_{i}) - 
(2\delta\alpha_i+2\delta\beta_it_i)\right)^2.
\end{eqnarray}
Let $\hat{\alpha}_i,\hat{\beta}_i$ minimize $Q(\alpha,\beta)$. The 
regression estimates of the instantaneous frequency
$f(t_i)$ and its rate of change $\dot{f}(t_i)$ 
are $\hat{\alpha}_i$ and $\hat{\beta}_i$ respectively. As these are 
obtained simultaneously, the $\dot{f}$ are on equal par with
$f$, rather than being their derivatives. Standard regression theory 
allows us to compute error boxes in these estimates. 
Note that the method 
amounts to an approximation
of the time-series $f(t)$ by a 
piecewise linear, and generally
discontinuous function.

The method may be optimized by modifying the present
$L^2$ norm in the error function to 
curve fitting, e.g.: different weights on
$f$ and $\dot{f}$. Initial experiments indicate that
the method can be implemented with adaptive window sizes, which may
improve its efficiency when the dynamic range in the
frequency $f(t)$ is large.

Because 
only the pattern of zero-crossings is measured, the algorithm
targets precisely the underlying gravitational wave frequency-luminosity 
relationship
of the source, as examplified in binary coalescence. 
The algorithm suppresses the amplitude of the signal, and hence
it is robust to size fluctuations in the signal.
Likewise,
reference to the underlying cosine functions 
is weak. In fact, any $2\pi$-function with two roots in 
each period can replace the cosine function in (1).

\section{Noise dependence}

The frequency dynamics algorithm has been tested in
simulated binary coalescence
using the Newtonian model of spiral in
($p=0$) with Gaussian noise and the instrumental 
noise from the 40m-LIGO test facility at Caltech.
The chirp extends over $t=[0,0.97)$, wherein the
number of periods at a given frequency $f(t)$
decreases from above 400 to below 200.
Figure 1 shows the results for both cases.

The algorithm obtains higher accuracy with the 40m-LIGO noise than 
with Gaussian noise when the variance is high. We believe that 
the reason for this lies in the high-degree of dependence in the  
noise from the LIGO prototype. 
The first-order autocorrelations 
from 40m-LIGO noise were of the order of 0.8. 
There were, on average, 15 zero-crossings per 100 recordings of
the pure noise process. This is much smaller than the 50 or so 
zero crossings obtained for white noise
of the same length. Thus the LIGO noise contributes  
far fewer ``false crossing'' due to noise as opposed to signal. This will 
be the case for all noise with low-frequency components in the spectrum. 
In such cases, our algorithms performs even better than it would with 
white noise. 
To make the last point more precise, consider a Gaussian process with 
a first order autocorrelation of $\rho$. If we observe a sequence of 
length $N$ from such a process the expected number of zero-crossings is 
\[\frac{(N-1)}{2}
\left(1-\frac{2}{\pi}\tan^{-1}\frac{\rho}{\sqrt{1-\rho^{2}}}\right)\]
For $\rho$ of 0.8, as in the LIGO case, we would expect approximately 20 
zero crossings per 100 recordings. Of course, the LIGO noise 
from the Caltech prototype is not pure Gaussian, but the principle is the same.

\section{{Comparison with matched filtering}}
Optimal application of 
matched filtering requires accurate
models of the system to
be detected and the noise of the detector. Both aspects of
prior knowledge are critical and must be described in a low
number of parameters, lest matched filtering becomes unstable.
Detection of gravitational waves from forementioned candidate
sources poses a number of uncertainties which may enter as model
imperfections (e.g., size of neutron stars\cite{BL:99}).
These imperfections can give rise to 
woefully wrong results 
when the true signal is not contained in the model for any
choice of its parameters. 
It becomes of interest, therefore, to compare the model-independence
aspect of the frequency dynamics algorithm with 
the sensitivity of matched filtering to these imperfections. 
The comparison would be complete with an additional on noise
type and intensity which, however, falls outside the scope of this
work. The comparison is made using
the post-Newtonian model in the 1PN approximation
as the true signal ($p=0.6038$), while applying matched filtering assuming
the Newtonian model given by $p=0$.
This mismatch of model to true signal is used here to
simulate other model imperfections, such as due to the 
finite size of the objects.

Thus, the matched filtering model is perfect in case of $p=0$,
and becomes imperfect as $p\ne0$. 
It is expected that matched filtering will outperform the
frequency dynamics algorithm for $p=0$, and vice-versa for
certain larger values of $p>0$.
The comparison is made quantitative using a mean absolute relative
error (MARE), expressed as a percentage, 
in the estimated frequencies by comparison with the
known signal input:
\begin{eqnarray}
\mbox{MARE}=
 \frac{100}{N}
 \sum_{1}^{N}
 \frac{ |\hat{f}(t_i)-f(t_i)| }{ f(t_i) }
\end{eqnarray} 
Using a small Monte Carlo study,
the comparison is made at three levels
of additive Gaussian white noise. Figure 2 shows the results
of this comparison study, where  
each reported MARE is the average of 200 simulations.

\section{Conclusions}

The trajectory in the $\dot{f}(f)$-plane of a burst of gravitational radiation
is expected to provide a signature of the source.
This is anticipated from the strong - possibly characteristic -
coupling between the
frequency and the luminosity of its gravitational wave emissions.
Binary coalescence, for example, shows the characteristic
chirp $\dot{f}\propto f^{\alpha+1/3}$ associated with
a luminosity ${\cal L}_{GW}\propto f^{\alpha}$, where $\alpha=10/3$
in general relativity in the point-particle limit.
Other burst sources of gravitational radiation 
may have different frequency dynamics
(e.g.: different $\alpha$'s).
We have described a nonparametric (model-free) frequency dynamics 
algorithm
to obtain an accurate plot of the trajectory in
the $\dot{f}(f)-$plane.
It is based on zero-crossings in the gravitational wave signal, to extract
these trajectories in the presence of noise.
The method is robust against non-Gaussian additive noise
and amplitude variations.
The method shows good results at appreciable chirp rates
and different types of noise. Even when the signal-to-noise
level is low (down to about 2 in our experiments), we can still recover
the distinctive curve of the frequency dynamics when $\dot{f}$ is
high. The estimation appears to be much more accurate for LIGO
noise than for Gaussian noise when the variance is high. We believe
that the reason for this lies in the high-degree of dependence
in the noise from the LIGO prototype, 
which suppresses ``false zero-crossings" generated by the noise.
This will be the case for all noise with low-frequency
components in the spectrum. In such cases, our algorithm performs
even better than it would with white noise.
The frequency dynamics algorithm has been compared with matched
filtering, and shown to be superior in the presence of
model imperfections. This is particularly relevant in searches
for unanticipated burst sources.

The present experiments indicate that the algorithm should perform
well in extracting the frequency dynamics in the gravitational waves.
Future experiments are desirable
at very low signal-to-noise levels, to study the
detection limit for transient sources with the
present algorithm.

\mbox{}\\
\mbox{}\\
{\bf Acknowledgements.} 
This research received partial
support from 
NASA grant NAG5-7012,
LIGO and an MIT Reed Award. The authors are grateful
to R. Weiss for constructive comments, 
and A. Lazzarini for kindly providing the LIGO data.

\newpage
\centerline{\bf Figure Captions}
\vskip1in
\mbox{}\\
{\bf Figure 1.} {The $\dot{f}(f)$-diagram of simulated
binary coalescence in the Newtonian approximation, 
extracted by the frequency dynamics
algorithm in the presence of noise at a signal-to-noise
ratio of about $4$ ($s=\sigma=0.25$).
The horizontal axis shows the frequency $f$ [Hz] and
the vertical axis its time rate of change $df$ [Hz/s].
The top window shows the results 
with Gaussian noise, and the lower window
shows the results with the 
instrumental noise from the 40m-LIGO 
test facility at Caltech.
The crosses in the right
windows denote horizontal and vertical errors 
obtained from the linear regression estimates.
The results indicate that the method improves
when the noise is correlated. 
At higher frequencies, the results are similar due to
a larger instantaneous signal-to-noise
ratio.
}
\mbox{}\\
\mbox{}\\
{\bf Figure 2.} 
{Experiments on model-independence
in terms of the normalized error MARE on the
frequency,
obtained by the frequency dynamics
algorithms (lines) and matched 
filtering (dashed) at
three different variances
$\sigma^2=0.01, 0.15, 0.25$.
The comparisons are
made for three different signals, 
parametrized by the value of $p$
(horizontal axis). Matched filtering
is performed with 
model assumption $p=0$, 
hence $p\ne0$ 
in the signal 
simulates a model imperfection.
The results show that
the frequency dynamics algorithm
outperforms
matched filtering in the presence of 
such appreciable model imperfections,
with MARE approaching a constant.
}
\newpage

\mbox{}\\
\vskip1in
\begin{center}
\epsfig{file=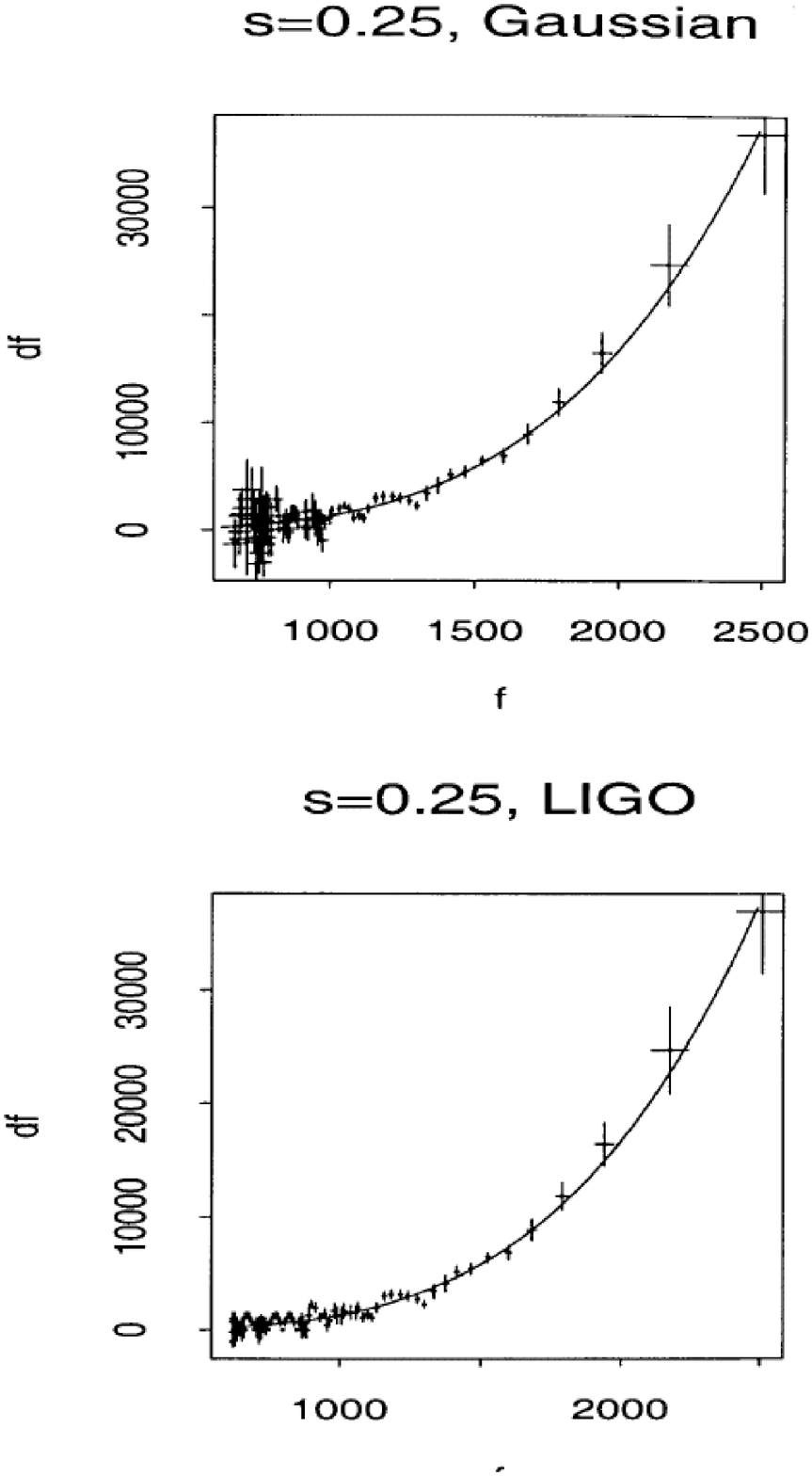,width=120mm,height=205.25mm}
\end{center}
{\bf Figure 1.}
\newpage
\mbox{}\\
\vskip1in
\begin{center}
\epsfig{file=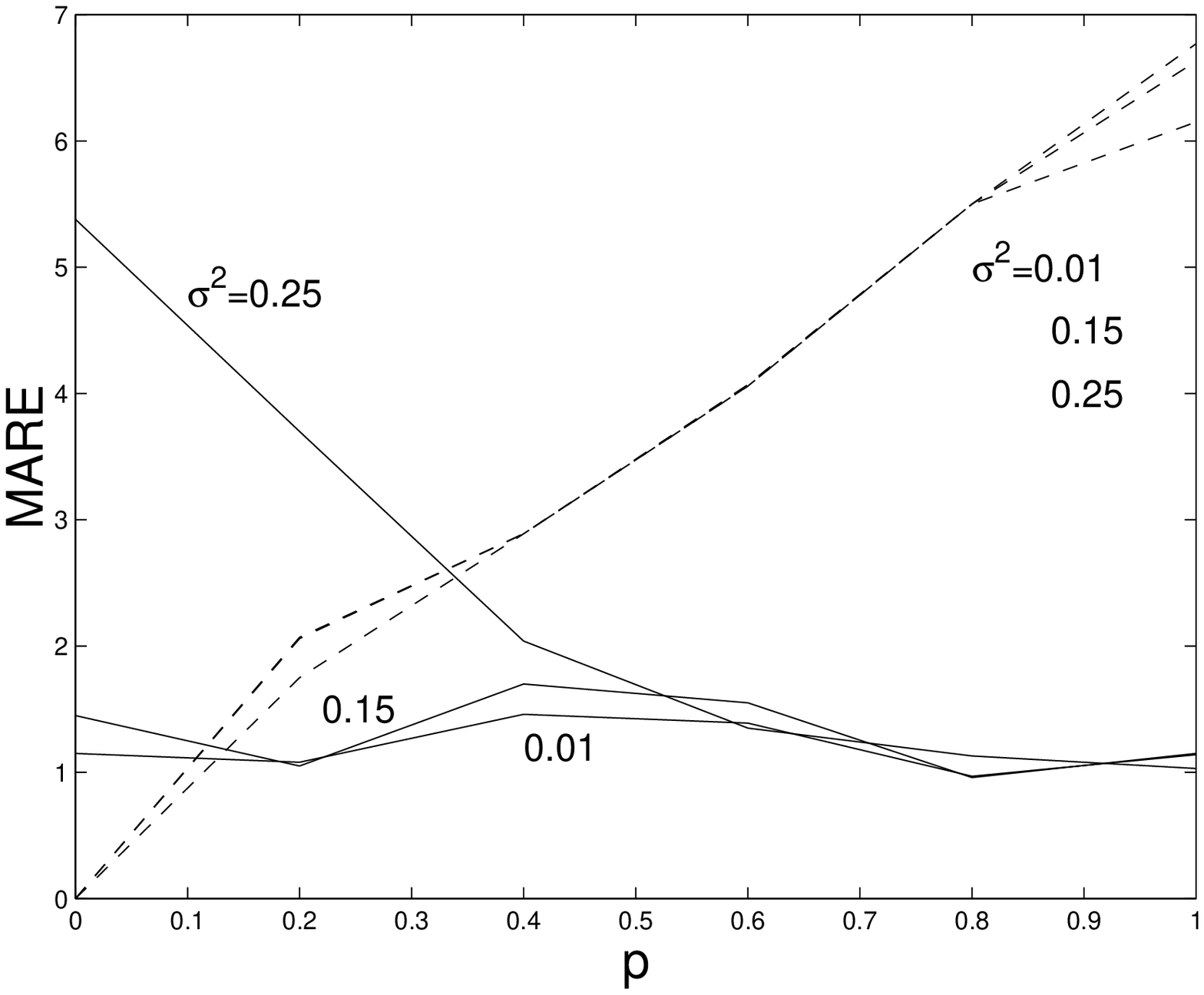,width=120mm,height=95.25mm}
\end{center}
{\bf Figure 2.}
\end{document}